\documentclass[10pt, conference]{IEEEtran}
\usepackage{algorithmicx}
\usepackage[ruled,vlined,linesnumbered]{algorithm2e}
\usepackage{hhline}
\usepackage{amsmath,mathtools}
\usepackage{amsfonts,amssymb}
\usepackage{mathrsfs}
\usepackage{caption}
\usepackage{enumitem}
\usepackage{multirow}
\usepackage{enumitem,color}
\usepackage{algpseudocode}

\begin{document}
%
\title{\huge \emph{ EdgeSlice}: Slicing Wireless Edge Computing Network with Decentralized Deep Reinforcement Learning}

\author{\IEEEauthorblockN{Qiang Liu, Tao Han, Ephraim Moges \vspace{-0.2in}}\\
\IEEEauthorblockA{Electrical and Computer Engineering Department,\\
The University of North Carolina at Charlotte, NC, United States\\
\{qliu12, tao.han, emoges\}@uncc.edu}\vspace{-0.3in}}

\maketitle

\begin{abstract}
5G and edge computing will serve various emerging use cases that have diverse requirements of multiple resources, e.g., radio, transportation, and computing. Network slicing is a promising technology for creating virtual networks that can be customized according to the requirements of different use cases. Provisioning network slices requires end-to-end resource orchestration which is challenging. In this paper, we design a decentralized resource orchestration system named EdgeSlice for dynamic end-to-end network slicing. EdgeSlice introduces a new decentralized deep reinforcement learning (D-DRL) method to efficiently orchestrate end-to-end resources. D-DRL is composed of a performance coordinator and multiple orchestration agents. The performance coordinator manages the resource orchestration policies in all the orchestration agents to ensure the service level agreement (SLA) of network slices. The orchestration agent learns the resource demands of network slices and orchestrates the resource allocation accordingly to optimize the performance of the slices under the constrained networking and computing resources. We design radio, transport and computing manager to enable dynamic configuration of end-to-end resources at runtime. We implement EdgeSlice on a prototype of the end-to-end wireless edge computing network with OpenAirInterface LTE network, OpenDayLight SDN switches, and CUDA GPU platform. The performance of EdgeSlice is evaluated through both experiments and trace-driven simulations. The evaluation results show that EdgeSlice achieves much improvement as compared to baseline in terms of performance, scalability, compatibility.
\end{abstract}


\section{Introduction}
\label{sec:introduction}
The emerging use cases and heterogeneous services, e.g., Internet of things (IoT), augmented/virtual reality (AR/VR), vehicle-to-everything (V2X), and mobile artificial intelligence, drive the development and research on the 5G mobile networks~\cite{agiwal2016next}.
Unlike the conventional services, these new services have highly diverse performance requirements such as bandwidth, delay, and reliability, which imposes a challenge for 5G to accommodate these services in terms of scalability, availability, and cost-efficiency~\cite{foukas2017network}.

Leveraging software defined networking (SDN) and network functions virtualization (NFV), network slicing is a promising technique to address this challenge~\cite{ordonez2017network}.
It enables multiple logical networks, i.e., network slices, run on top of a common physical network infrastructure~\cite{afolabi2018network}.
Network slices can be individually customized to meet various performance requirements of different network services and use cases.
For example, a slice can be customized to carry IoT services that require massive connections but low data rates.
At the same time, another slice may be instantiated to support delay-sensitive services, e.g., mobile augmented reality and vehicle-to-vehicle communication.
Thus, network slicing creates new network management and operation patterns and improves network performance for both the network operator and service providers in terms of network revenue, quality of service, and service autonomy.

Network operator is required to provide the performance and functional isolation to network slices~\cite{foukas2017orion}.
The performance isolation ensures that the performance of a network slice is not affected by the operation of the other network slices.
The functional isolation allows slice tenants to customize their slice's functions and resource management~\cite{rost2017network}.
However, the isolation among network slices reduces the multiplexing efficiency and thus degrades the system performance~\cite{marquez2018should}.
It is observed that the multiplexing efficiency improves when the network resources are shared in a small time scale~\cite{marquez2018should}.
This observation advocates the dynamic network slicing which can dynamically change the resource allocation in network slices according to their actual needs.

\begin{figure}[!t]
	\centering
	\includegraphics[width=0.9\linewidth]{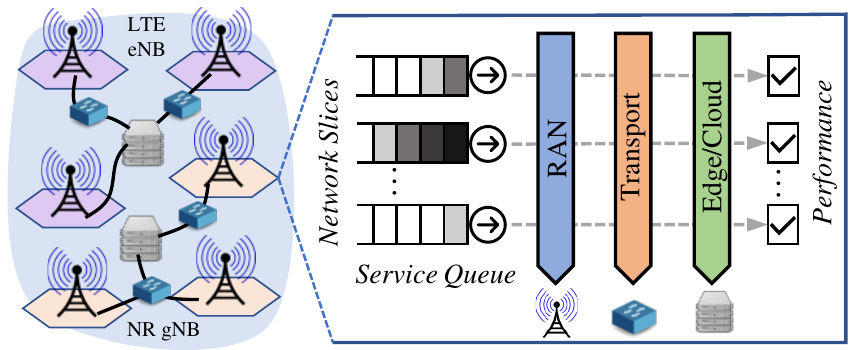}
	\caption{\small An illustration of end-to-end network slicing in wireless edge computing network composed of distributed network infrastructures such as BSs and servers. Network slices require end-to-end resources from multiple technical domains, e.g., radio access network, transport network and edge/cloud computing platform, to serve their slice users.}
	\label{fig:system_model}
	\vspace{-0.1in}
\end{figure}

Dynamic network slicing, as illustrated in Fig.~\ref{fig:system_model}, faces two research challenges. First, it is almost impossible to obtain the exact correlation between the resources and performance of network slices. A network slice usually requires resources from multiple technical domains such as radio access network, transport network, and edge/cloud. There are very complex tradeoffs among these resources and slice performance. For example, a short delay in the radio access network can be compensated by accelerated computation in the edge/cloud servers. Therefore, there lacks a closed-form mathematical expression that models the correlation between the resources and performance of network slices. The existing works on multi-resource allocation usually assume that multiple resources are allocated following a certain ratio, e.g., 1 unit radio spectrum : 2 unit computing resources, which is not efficient~\cite{caballero2019network, halabian2019distributed}. The second challenge is that the spatial diversity of mobile traffic requests the resources of network slices to be properly distributed among base stations and edge/cloud servers in different geographic locations. This further complicates the dynamic network slicing problem.

In this paper, we design \emph{EdgeSlice}, a decentralized resource orchestration system that automates dynamic end-to-end network slicing in wireless edge computing networks. EdgeSlice introduces a novel decentralized deep reinforcement learning (D-DRL) method to efficiently orchestrate end-to-end networking and computing resources. With the D-DRL methods, the resource orchestration is carried out by a central performance coordinator and multiple decentralized orchestration agents. The orchestration agents rely on DRL to learn optimal resource orchestration policy, and the central performance coordinator coordinates the resource orchestration in the agents to ensure the service level agreements (SLAs) of network slices. To realize EdgeSlice, we also develop new radio, transport, and computing resource manager that can manage the resources at runtime according to the resource orchestration actions and instantiate network slices.


The contributions of this paper are summarized as follows:
\begin{itemize}
    \item We design and implement EdgeSlice which is a decentralized resource orchestration system for dynamic network slicing in wireless edge computing networks. EdgeSlice automates dynamic network slicing leveraging a novel decentralized deep reinforcement learning (D-DRL) method.
    \item We design a new D-DRL method to automate the end-to-end resource orchestration with high efficiency. The D-DRL method is composed of a performance coordinator and multiple orchestration agents. The orchestration agent can learn the optimal resource orchestration policy under the coordination of the performance coordinator. 
    \item We develop radio, transport and computing manager which are integrated with existing platforms: OpenAirInterface (OAI) in radio access network, OpenDayLight (ODL) in transport network, and CUDA GPU in edge/cloud servers. These managers enable the dynamic configuration of end-to-end resources at runtime in the EdgeSlice system.  
    \item We build an experiential prototype and implement the EdgeSlice system. We evaluate the performance of the EdgeSlice system through both experiments using the prototype system and trace-driven network simulations. 
\end{itemize}


\section{EdgeSlice Overview}
\label{sec:edgeslice}
\begin{figure*}[!t]
\centering
\includegraphics[width=6in]{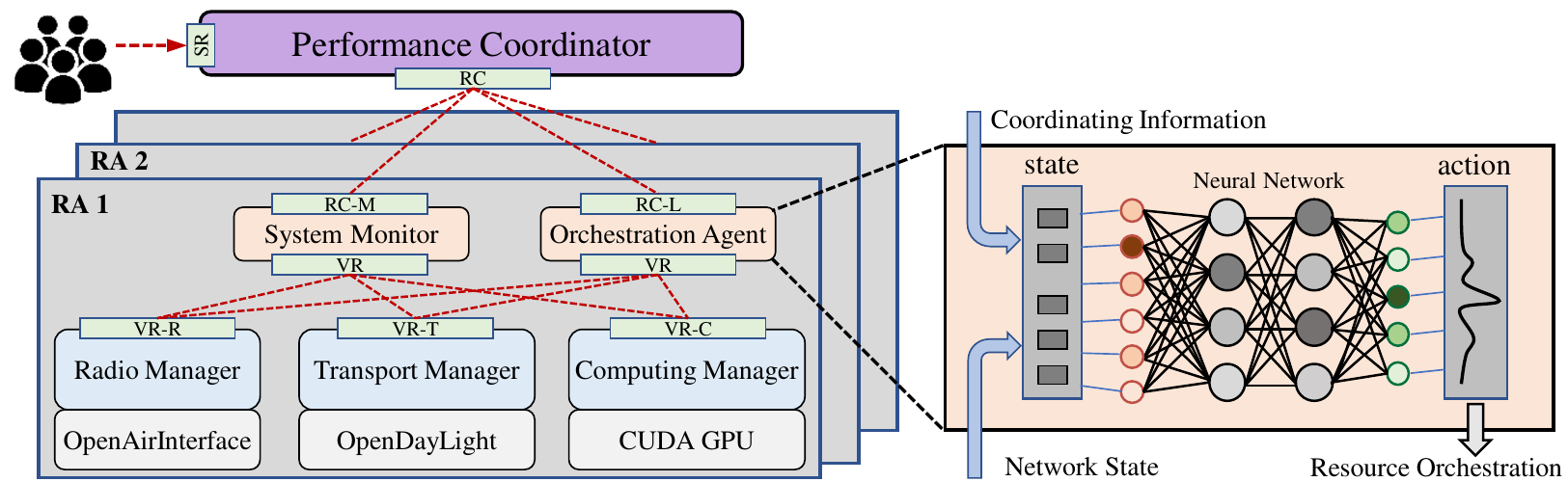}
\caption{The EdgeSlice System.}
\vspace{-0.1in}
\label{fig:system_design}
\end{figure*}

EdgeSlice automates dynamic network slicing in wireless edge computing networks through decentralized deep reinforcement learning. Fig.~\ref{fig:system_design} outlines the design of the EdgeSlice system. 
To automate the network slicing process, EdgeSlice leverages machine learning, i.e., deep reinforcement learning, to learn end-to-end resource demands of network slices and then orchestrates the resource allocations to network slices accordingly. Owing to the temporal and spatial dynamics of the slice traffic and the complex tradeoffs between the performance of network slices and the resource orchestration, it is inefficient to use a centralized learning agent to orchestrate resource allocations to network slices. Besides, a centralized learning agent needs to obtain network performance data from all the network nodes, which introduces excessive communication overhead and delay. Toward this end, EdgeSlice introduces a new decentralized deep reinforcement learning method for network slicing in wireless edge computing networks. 

We define a resource autonomy (RA) as a set of network infrastructures such as BSs and edge servers in a geographic area, and thus the network can be partitioned into multiple RAs.
An orchestration agent is designed based on deep reinforcement learning to manage multi-domain resources in each RA and operates on a short timescale, e.g., seconds, to enable dynamic network slicing. The orchestration agent (detailed in Sec.~\ref{sec:orchestration_agent}) can track the network state (queue length, traffic), learn the resource orchestration policy from experience and orchestrate resources to slices autonomously.

A centralized performance coordinator is designed to coordinate the resource orchestration in all the RAs and optimizes the performance of the network on a much larger timescale. Meanwhile, the performance coordinator ensures that all the constraints related to the resource orchestration, e.g., SLAs and system capacity, are satisfied (detailed in Sec.~\ref{sec:performance_coordinator}). The performance coordinator only exchanges slight coordinating information with orchestration agents, which substantially decreases the communication overheads.

To realize EdgeSlice, resource managers, i.e., middleware, are developed to manage resources in radio access network, transport network, and edge computing servers at runtime according to the resource orchestration decision made by orchestration agents (detailed in Sec.~\ref{sec:control_plane}).

\begin{table}[!t]
    \footnotesize
	\centering
	\begin{tabular}{|c|c||c|c|}
        \hline
        \textbf{entity}     &  \textbf{symbol}          &  \textbf{entity}              &  \textbf{symbol} \\ 
        \hline
        network slice       &      $i$                  &     resource autonomy (RA)    & $j$ \\ 
        network resource    &      $k$                  &    time interval                  & $t$ \\ 
        slice queue length  &      $l$                  &    time period                & $\mathcal{T}$ \\ 
        slice performance   &      $\mathbf{U}$         &     resource orchestration    & ${x}$ \\ 
        min. performance    &   $\mathbf{U}^{\min}$     &     total resource            & $R^{tot}$ \\ 
        auxiliary variable  &   $z$                     &     dual variable            & $y$ \\ 
        \hline
	\end{tabular}
	\caption{ Notations throughout Sec.\ref{sec:framework}}\label{notations}
\vspace{-0.2in}
\end{table}


\section{System Model and Problem Statement}
\label{sec:framework}
To design the EdgeSlice system, we first mathematically model the wireless edge computing network and formalize the statement of end-to-end resource orchestration problem. 

\subsection{System Model}
\label{sec:system_mdel}
We consider an end-to-end wireless edge computing network which is composed of a radio access network (RAN) with multiple base stations (BSs), edge/cloud computing servers, and a transport network connecting the RAN and computing servers.
As shown in Fig.~\ref{fig:system_model}, there are multiple network slices that request end-to-end resources in every RA, in order to enable seamless service coverage and support their users mobility.
In each RA, network slices have service queues that buffer the arrival traffic of their slice users.
We consider the network is time-slotted, and network operator can observe the performance\footnote{Network slices could have various metrics on evaluating their performances, e.g., latency, throughput, queue status.} of network slices and dynamically change its resource orchestration with a minimum $t$ time interval.

Let $\mathcal{I}$, $\mathcal{J}$ and $\mathcal{K}$ be the sets of network slices, RAs and network resources, respectively.
Denote $\mathbf{x}_{i,j}^{(t)} = [ x_{i,j,k}^{(t)}|\forall k\in\mathcal{K}]$ where $x_{i,j,k}^{(t)}$ is the $k$th resource allocated to the $i$th slice on the $j$th RA and $\mathbf{U}_{i,j}^{(t)}$ is the performance of network slice.

\subsection{Problem Statement}
The objective of network slicing is to maximize the performance of network slices in the system, and the objective of the network slicing can be expressed as 
$\max \limits_{\{ \mathbf{x}_{i,j}^{(t)}\}} \; \lim\limits_{\tau \to \infty} \frac{1}{\tau} \sum\limits_{t = 0}^{\tau} {\sum\limits_{i \in \mathcal{I}} \sum\limits_{j \in \mathcal{J}} \mathbf{U}_{i,j}^{(t)}}.$
As $\tau \to \infty$, the problem is an infinite time horizon stochastic programming problem. A general method to solve the problem is to transform it into a problem within a finite time period $\mathcal{T}$, e.g., a day~\cite{kall1994stochastic,salvat2018overbooking}. Hence, the resource orchestration problem is formulated as 
\vspace{-0.05in}
\begin{equation}
    \label{prob1}
	\begin{array}{*{20}{l}}
		{\mathscr{P}_0:}&{\max \limits_{\{ \mathbf{x}_{i,j} \ge 0\}}}&{  \sum\limits_{t \in \mathcal{T}} {\sum\limits_{i \in \mathcal{I}} \sum\limits_{j \in \mathcal{J}} \mathbf{U}_{i,j}^{(t)}} }\\
		{}&{s.t.}&{(\ref{const1}), (\ref{const2})}.\\
	\end{array}
\vspace{-0.05in}
\end{equation}

In the context of network slicing, the resource orchestration problem subjects to two practical constraints.
The first constraint is that the network-wide performance of a network slice should meet the SLA made between the slice tenant and network operator. Denote $\mathbf{U}_{i}^{\min}$ as the minimum performance requirement of the $i$th slice according to the SLA.
Thus, the performance constraint can be written as
\vspace{-0.05in}
\begin{align}
\label{const1}
	&{\; \sum\nolimits_{t \in \mathcal{T}}{\sum\nolimits_{j \in \mathcal{J}} \mathbf{U}_{i,j}^{(t)}} \ge \mathbf{U}_{i}^{\min},}&{ \forall i \in \mathcal{I}}.
\vspace{-0.05in}
\end{align}

The second constraint is that the resources in each RA are limited.
Denote $R^{tot}_j=[r^{tot}_{j,k}|\forall k \in\mathcal{K}]$ as the total amount of each resource in the $j$th RA.
Then, the resource allocated to network slices in the $j$th RA should be less than $R^{tot}_j$, and the constraint can be expressed as 
\vspace{-0.05in}
\begin{align}
\label{const2}
	&{\;\sum\nolimits_{i \in \mathcal{I}} \mathbf{x}_{i,j}^{(t)} \le R^{tot}_j, }&{\forall j\in \mathcal{J}, t \in \mathcal{T}}.
\vspace{-0.05in}
\end{align}

The difficulties in solving problem $\mathscr{P}_0$ are two-fold. First, the problem involves the end-to-end resource orchestration to network slices within each RA and the performance coordination across all RAs to maintain network-wide performance of network slices. The coupling between the intra-RA and inter-RAs resource management highly complicates the problem. Second, due to the varying network dynamics and the diversity of resource demands of network slices, the slice performance becomes a complex stochastic function. In real systems, it is almost impossible to derive an accurate mathematical model for such correlation~\cite{mao2019learning}.
Moreover, the resource orchestration in the network slicing system exhibits \emph{Markovian} on serving slice users where a resource orchestration policy affects not only the current but also further network state, e.g., service queues.

\section{EdgeSlice Design: Coordinator and Agents}
\label{sec:design}
In this section, we present the design of performance coordinator and orchestration agents in the EdgeSlice system.

\subsection{Performance Coordinator}
\label{sec:performance_coordinator}
Since the performance of a network slice depends on the resource orchestration in multiple RAs, the central performance coordinator is designed to coordinate the resource orchestration among RAs and thus optimizes the performance of the network slices. 
To design the performance coordinator, we transform problem $\mathscr{P}_0$ by introducing auxiliary variables $\mathcal{Z} =\{ z_{i,j}, \forall i \in \mathcal{I}, j \in \mathcal{J}\}$ where
\vspace{-0.07in}
\begin{align}
    \label{const4}
    &{z_{i,j} =  \sum\nolimits_{t \in \mathcal{T}}  \mathbf{U}_{i,j}^{(t)},} &{ \forall i \in \mathcal{I}, j \in \mathcal{J}}.
\vspace{-0.07in}
\end{align}
Then, the constraint (\ref{const1}) are equivalent to 
\vspace{-0.07in}
\begin{align}
    \label{const5}
    &{ {\sum\nolimits_{j \in \mathcal{J}} z_{i,j}} \ge \mathbf{U}_{i}^{\min},} & { \forall i \in \mathcal{I}}.
\vspace{-0.07in}
\end{align}
Hence, problem $\mathscr{P}_0$ is equivalently transformed to
\vspace{-0.07in}
\begin{equation}
	\label{prb:transformed_problem}
	\begin{array}{*{20}{l}}
		{\mathscr{P}_1:}&{\max \limits_{\{ \mathbf{x}_{i,j} \ge 0,z_{i,j}\}}}&{ \sum\limits_{t \in \mathcal{T}} {\sum\limits_{i \in \mathcal{I}} \sum\limits_{j \in \mathcal{J}} \mathbf{U}_{i,j}^{(t)}} }\\
		{}&{s.t.}&{(\ref{const2}), (\ref{const4}), (\ref{const5})}.\\
	\end{array}
\vspace{-0.07in}
\end{equation}
Problem $\mathscr{P}_1$ has two sets of variables, $\mathcal{X}$ and $\mathcal{Z}$ which are coupled by constraint (\ref{const4}).
Next, we derive augmented Lagrangian of problem $\mathscr{P}_1$ as
\vspace{-0.07in}
\begin{equation}
	\mathcal{L}_y =  {\sum\limits_{i \in {\mathcal{I}}} {\sum\limits_{j \in {\mathcal{J}}}{\left(  \sum\limits_{t \in \mathcal{T}} {{\bf{U}}_{i,j}^{(t)} - \frac{\rho }{2}\| { \sum\limits_{t\in \mathcal{T}}{\bf{U}}_{i,j}^{(t)}  - z_{i,j} + y_{i,j}} \|_2^2} \right)} }},
\vspace{-0.05in}
\end{equation}
where $\rho \geq 0$ is a positive constant, and $\mathcal{Y} =\{ y_{i,j}, \forall i \in \mathcal{I}, j \in \mathcal{J}\}$ is the scaled dual variables.
Here, the augmented Lagrangian incorporates the constraint (\ref{const4}) which couples variables $\mathcal{Z}$ and $\mathcal{X}$. 

According to the alternating direction method of multipliers (ADMM) method~\cite{boyd2011distributed}, problem $\mathscr{P}_1$ is solved by iteratively solving the following problems:
\vspace{-0.07in}
\begin{align} 
	&{\mathbf{x}_{i,j}= \arg \max_{\mathbf{x}_{i,j} \in (\ref{const2})} \mathcal{L}_y (\mathbf{x}_{i,j},z_{i,j},y_{i,j})}, \label{x-update}\\
	&{z_{i,j}= \arg \max_{z_{i,j} \in (\ref{const5})} \mathcal{L}_y (\mathbf{x}_{i,j},z_{i,j},y_{i,j})}, \label{z-update}\\
	&{y_{i,j}= y_{i,j} +(  \sum\nolimits_{t \in \mathcal{T}} \mathbf{U}_{i,j}^{(t)} - z_{i,j})} \label{y-update},
\vspace{-0.07in}
\end{align}
where problem in Eq.~\ref{x-update} focuses on the resource orchestration. Problem in Eq.~\ref{y-update} and Eq.~\ref{z-update} update auxiliary and dual variables, respectively, which require all the resource orchestrations in the system.  

Therefore, we design the performance coordinator to solve the problem in Eq.~\ref{z-update} and Eq.~\ref{y-update} based on the resource orchestration and slice performance collected from orchestration agents in the system.
Since $\mathcal{X} $ and $\mathcal{Z}$ are obtained, the problem in Eq.~\ref{z-update} is equivalent to
\vspace{-0.07in}
\begin{equation}
	\begin{array}{*{20}{l}}
		{\mathscr{P}_2:}&{\min \limits_{\{ z_{i,j}\}}}&{  \sum\limits_{i \in \mathcal{I}} { \sum\limits_{j \in \mathcal{J}} \|  {\sum\limits_{t \in \mathcal{T}}{\bf{U}}_{i,j}^{(t)}  - z_{i,j} + y_{i,j}} \|_2^2}}\\
		{}&{s.t.}&{(\ref{const5})}.\\
	\end{array}
\vspace{-0.07in}
\end{equation}
This problem is a standard quadratic programming problem which can be solved by using convex optimization tools, e.g., CVX~\cite{Convex2004Boyd}.
By solving the problem, the performance coordinator obtains auxiliary variables $\mathcal{Z}$ and then updates dual variables $\mathcal{Y}$ according to Eq.~\ref{y-update}.
We define the auxiliary variables $\mathcal{Z}$ and the dual variables $\mathcal{Y}$ as the coordinating information between the performance coordinator and orchestration agents. 

\subsection{Orchestration Agent}
\label{sec:orchestration_agent}
The orchestration agents are designed to orchestrate the end-to-end resources for network slices under the supervision of the performance coordinator, i.e., solving the problem in Eq.~\ref{x-update}.
Since the constraint of the problem only restricts the resource orchestration within a RA, it can be solved individually within each RA, i.e. decentralized. Hence, we rewrite the problem in Eq.~\ref{x-update} within the $j$th RA as
\vspace{-0.07in}
\begin{equation}
	\begin{array}{*{20}{l}}
		{\mathscr{P}_3:}&{\max \limits_{\{ \mathbf{x}_{i,j} \ge 0 \}}}&{{\sum\limits_{i \in {\mathcal{ I}}}  {\sum\limits_{t \in \mathcal{T}}{\bf{U}}_{i,j}^{(t)} } }}\\
		{}&{}&{- \frac{\rho }{2}{\sum\limits_{i \in {\mathcal{ I}}}\| {\sum\limits_{t \in \mathcal{T}}{\bf{U}}_{i,j}^{(t)}  - z_{i,j} + y_{i,j}} \|_2^2}}\\
		{}&{s.t.}&{(\ref{const2})}.\\
	\end{array}
	\label{prob:slave_problem}
\vspace{-0.07in}
\end{equation}
The major challenge of solving the above problem is that the slice performance is very complex and without a closed-form mathematical model because of the varying network dynamic and the complicated end-to-end resource demands of network slices.
Moreover, the current resource orchestration impacts both slice users in service queues and further network state.
To address this challenge, we resort to deep reinforcement learning (DRL) techniques that enable model-free machine learning~\cite{lillicrap2015continuous} when designing orchestration agents.

We consider a general reinforcement learning setting where an agent interacts with an environment in discrete decision epochs.
At each decision epoch $t$, the agent observes a state $\mathbf{s}_{t}$, takes an action $\mathbf{a}_t$, e.g., resource orchestration, based on its policy $\mu(\mathbf{s})$, and receives a reward $r(\mathbf{s}_t, \mathbf{a}_t)$.
Then, the environment transits to the next state $\mathbf{s}_{t+1}$, e.g., queue status changes, based on the action taken by the agent.
The objective is to find the optimal policy $\mu^*(\mathbf{s})$ mapping states to actions, that maximizes the discounted cumulative reward $\sum\nolimits_{t=0}^{\infty} \gamma^t r(\mathbf{s}_t, \mathbf{a}_t)$. Here, $\gamma \in [0, 1]$ is a discounting factor. 

Although DRL techniques have been extensively studied in many areas such as robotic control~\cite{mnih2015human}, traffic control~\cite{xu2018experience}, and chess games~\cite{silver2018general}, the existing DRL models are not appropriate to solve problem $\mathscr{P}_3$ for two reasons. First, most of the DRL models are designed to solve constraint-free problems~\cite{xu2018experience, chen2018auto}. However, the problem consists of multiple linear constraints. Second, the existing DRL models are unable to adjust their policies based on coordinating information from an external control~\cite{mao2016resource}. However, to maintain the network-wide performance of network slices, the agent in EdgeSlice needs to orchestrate resources according to the coordinating information derived from the coordinator.


\subsubsection{Design of Agents}
Therefore, we design a new DRL model with customized state space, action space and reward function. In the DRL model, the constraint (\ref{const2}) are re-weighted and incorporated into its reward function so that the reward is affected by whether the constraints are satisfied or not. The coordinating information is augmented into state space to allow external control from the coordinator. 


\textbf{State Space}:
The state is concatenated by two parts. The first part is $[l_{j}^{(t)}, \forall i \in \mathcal{I}]$ which represents the current network state, i.e., queue status of network slices.
The second part is $[z_{i,j}-y_{i,j}, \forall i \in \mathcal{I}]$ which is the coordinating information from the coordinator. Thus, the state in the $j$th RA at time interval $t$ can be expressed as
\vspace{-0.05in}
\begin{equation}
    \mathbf{s}_t = \left[l_{j}^{(t)}, z_{i,j}-y_{i,j}, \forall i \in \mathcal{I}\right].
\vspace{-0.05in}
\end{equation}

\textbf{Action Space}:
The action at time interval $t$ is defined as the resource allocations to network slices in the RA:
\vspace{-0.05in}
\begin{equation}
    \mathbf{a}_t = \left[  \mathbf{x}_{i,j}^{(t)}, \forall i \in \mathcal{I} \right].
\vspace{-0.05in}
\end{equation}

\textbf{Reward}:
The reward at time interval $t$ is defined as
\vspace{-0.05in}
\begin{align}
\label{eq:reward_function}
    r(\mathbf{s}_t, \mathbf{a}_t) =& \sum\limits_{i \in {\mathcal{I}}} \left( {{\bf{U}}_{i,j}^{(t)} - \frac{\rho }{2} \| {  {\bf{U}}_{i,j}^{(t)}  - \frac{1}{\mathcal{T}}( z_{i,j} + y_{i,j})} \|_2^2} \right)  \\ \nonumber
    &{- \beta \sum\nolimits_{j \in \mathcal{J}}  \left[ {\sum\nolimits_{i \in \mathcal{I}} \mathbf{x}_{i,j}^{(t)} - R^{tot}_j} \right]^+}, 
\vspace{-0.05in}
\end{align}
where $[x]^+ = \max \left(  0, x \right)$, and $\beta$ is a positive constant. 
Here, we approximate the objective function of problem $\mathscr{P}_3$ with identical sub-objective functions in the time domain.
Moreover, we incorporate the constraints (\ref{const2}) into the sub-objective functions with reward shaping technique~\cite{griffith2013policy}.
Therefore, there will be a penalty added into the reward if the constraints are violated.

\subsubsection{Training of Agents}
\begin{figure}[!t]
	\centering
	\includegraphics[width=3.3in]{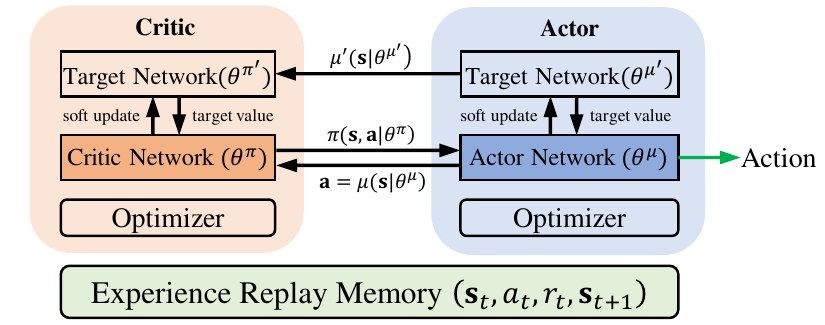}
	\caption{\small The DDPG architecture.}
	\label{fig:DDPG}
	\vspace{-0.2in}
\end{figure}

We follow deep deterministic policy gradient (DDPG), a state-of-the-art reinforcement learning technique that is capable of handling continuous and high-dimensional action spaces~\cite{lillicrap2015continuous}, to train our orchestration agents.
As shown in Fig.~\ref{fig:DDPG}, DDPG integrates deep Q-network (DQN) \cite{mnih2015human} and actor-critic method~\cite{konda2000actor}, and maintains a parameterized actor $\mu(\mathbf{s}_t | \theta^\mu)$ and a parameterized critic $\pi(\mathbf{s}_t,\mathbf{a}_t | \theta^\pi)$.
The critic estimates the value function of state-action pairs, and the actor specifies the current policy by mapping a state to a specific action.

The critic is implemented using a DQN. We define the value function $ Q^\pi(\mathbf{s}_t, \mathbf{a}_t)$ as the expected discounted cumulative reward when the agent starts with the state-action pair $(\mathbf{s}_t, \mathbf{a}_t)$ at decision epoch $t$ and then acts according to the policy $\pi$. Then, the value function can be expressed as
$Q^\pi(\mathbf{s}_t, \mathbf{a}_t) = \mathop{\mathbb{E}}\nolimits_{ \pi} {\left[ R_t \right]},$
where $R_t = \sum\nolimits_{k=t}^T \gamma^{(k-t)} r(\mathbf{s}_k, \mathbf{a}_k) $.
Based on the Bellman equation~\cite{bellman1966dynamic}, the optimal value function is
$Q^*(\mathbf{s}_t, \mathbf{a}_t) = r(\mathbf{s}_t, \mathbf{a}_t) + \gamma \max\limits_{\mathbf{a}_{t+1}}Q^*(\mathbf{s}_{t+1}, \mathbf{a}_{t+1})$.

To obtain the optimal policy, DQN is trained by minimizing the mean-squared Bellman error (MSBE)
\vspace{-0.05in}
\begin{equation}  
    L(\theta^\pi) = \mathop{\mathbb{E}}{\left[ {\left( g_t - Q(\mathbf{s}_t, \mathbf{a}_t | \theta^\pi)\right)}^2\right]},
\vspace{-0.05in}
\end{equation}
where $\theta^\pi$ are parameters of the critic network, and $\mathcal{D}$ is a replay memory.
$g_t$ is the target value estimated by a target network, and can be expressed as 
\vspace{-0.05in}
\begin{equation}
    g_t = r(\mathbf{s}_t, \mathbf{a}_t) + \gamma \max\nolimits_{\mathbf{a}_{t+1}}Q(\mathbf{s}_{t+1}, \mu(\mathbf{s}_{t+1}|\theta^{\mu'} )| \theta^{\pi'}),
\vspace{-0.05in}
\end{equation}
where $\theta^{\pi'}$ are parameters of the target network.
The target network has the same architecture as the critic network, and its parameters $\theta^{\pi'}$ are slowly updated to track that of the critic network.

The actor is implemented using another DQN which learns a deterministic policy $\mu(\mathbf{s}_t | \theta^\mu)$ to maximize the cumulative reward of the actor, i.e., $J = \mathop{\mathbb{E}}\nolimits_{ \mu} {\left[ R_t \right]}$. Since the action space is continuous, the value function is assumed to be differentiable with respect to the action.
Thus, the actor network can be trained by applying the chain rule to the expected cumulative reward with respect to the actor parameters $\theta^\mu$:
\vspace{-0.05in}
\begin{align}
    \nabla_{\theta^\mu} J &\approx \mathbb{E}_{ }\left[\nabla_{\theta^\mu} Q(\mathbf{s},\mathbf{a} | \theta^\pi) |_{\mathbf{s}=\mathbf{s}_t,\mathbf{a}=\mu(\mathbf{s}_t|\theta^\mu)} \right]\\ \nonumber
    &= \mathbb{E}_{ }\left[\nabla_{\mathbf{a}} Q(\mathbf{s},\mathbf{a} | \theta^\pi) |_{\mathbf{s}=\mathbf{s}_t,\mathbf{a}=\mu(\mathbf{s}_t)}  \cdot \nabla_{\theta^\mu} \mu (\mathbf{s}|\theta^\mu)|_{\mathbf{s}=\mathbf{s}_t} \right]. 
\vspace{-0.05in}
\end{align}



\begin{algorithm}[!t]
	\caption{The EdgeSlice Resource Orchestration}\label{alg:proposed}

	\KwIn{$\mathbf{U}^{\min}_{i}$, $\forall i \in \mathcal{I}$; $R^{tot}_j,\forall i \in \mathcal{I}$; $\rho$, $\beta$. }
	\KwOut{$\mathcal{X, Z, Y}$.}
    Initialize $\mathcal{Z}$ and $\mathcal{Y}$\;
	\While{True}
	{
    	$/**\;optimize\;\mathcal{X}\; in\; each\; agent\; **/$\;
    	\For{$j \in \mathcal{J} \; (decentralized)$}
    	{
    	    $\mathbf{x}_{i,j}^{(t)}$, $\forall i\in\mathcal{I},t \in \mathcal{T} \gets$ the $i$th orchestration agent\;
    	    $\mathbf{U}_{i,j}^{(t)}$, $\forall i\in\mathcal{I},t \in \mathcal{T} \gets $ the $i$th slice performance\;
    	   
    	}
    	
    	$/**\;update\;\mathcal{Z}\; in\; the\; coordinator\; **/$\;
    	$z_{i,j} \gets \arg \max\limits_{z_{i,j} \in (\ref{const5})} \mathcal{L}_y (\mathbf{x}_{i,j},z_{i,j},y_{i,j})$\;
    	$/**\;update\; \mathcal{Y}\; in\; the\; coordinator\; **/$\;
    	$y_{i,j} \gets y_{i,j} +( \sum\nolimits_{t \in \mathcal{T}} \mathbf{U}_{i,j}^{(t)} - z_{i,j})$\;
    	
    	$/**\;if\; algorithm\; convergence\;**/$\;
    	\If{convergence}
    	{   \Return{$\mathcal{X, Z, Y}$}\;
    	}
	}
\end{algorithm}

\subsection{The Workflow of EdgeSlice}
The workflow of the EdgeSlice system is summarized in Alg.~\ref{alg:proposed}.
The resource orchestration starts by initializing the coordinating information, i.e., $\mathcal{Z}$ and $\mathcal{Y}$.
The orchestration agent in each RA orchestrates resources to network slices based on its parameterized policy under the coordinating information for time intervals in $\mathcal{T}$.
At the end of a time period $\mathcal{T}$, the orchestration agent collects the performance of network slices $\textbf{U}$.
Given $\mathcal{X}$ and $\textbf{U}$, the performance coordinator generates the coordinating information ($\mathcal{Y}$ and $\mathcal{Z}$), which are fed back to orchestration agents in all RAs.
It continues until the convergence of the resource orchestration.



\section{EdgeSlice Design: Resource Manager}
\label{sec:control_plane}
In this section, we design radio, transport, and computing manager that allocates the resources orchestrated by agents to network slices at runtime, as shown in Fig.~\ref{fig:system_design}. These managers are integrated with OpenAirInterface (OAI), OpenDayLight (ODL), and CUDA GPU computing platform to enable dynamic configuration of resources in radio access network, transport network, and edge/cloud computing, respectively.

\subsection{Radio Manager}
The radio manager is designed to work with OpenAirInterface (OAI) to allocate radio resources to slice users in both uplink (UL) and downlink (DL) radio access network.
In EdgeSlice, the total radio resources (bandwidth) can be used by a network slice is determined by the orchestration agent.
Once a network slice obtains its radio resources, it allocates these resources to its users.
As a result, the allocated radio resources of all slice users are known by the radio manager.
Hence, the radio manager should schedule users according to their allocated resources at runtime, which is not supported by vanilla OAI.

We fulfill such functionality by developing a new user scheduling method in the MAC layer to manage physical resource blocks (PRBs) in PUSCH/PDSCH.
We schedule the slice users consecutively and map their radio resources to PRBs.
The users without any radio resources will not be scheduled.
To support the information exchange between the orchestration agent and the radio manager at runtime, we develop the VR-R (virtual resource - radio) and VR (virtual recourse) interfaces in the radio manager and orchestration agent, respectively. 
The association between a mobile user and a network slice is identified by the user's international mobile subscriber identity (IMSI).
The IMSI information is extracted from the S1AP message sent from the base station to mobile management entity (MME).
The information extraction does not need any modification on the mobile user's side.

\subsection{Transport Manager}
Taking advantaging of the separation of data and control plane in SDN switches, we allocate the bandwidth of links between RAN and edge/cloud computing servers with an OpenDayLight~\cite{medved2014opendaylight} controller through OpenFlow (Southbound API) and RESTful (Northbound API)~\cite{mckeown2008openflow}.
The OpenFlow protocol currently supports user bandwidth modification with \emph{meters}. However, these meters and their attached flows should be deleted and reinitialize if the user bandwidth needs to be changed.
As a result, when changing the user bandwidth allocation at runtime, the switch network is broken during the deletion-creation interval~\cite{openflow_ref}.

To enable dynamic modification of bandwidth while keeping the switches network alive, we create a new configuration that parallels with the current one when a new user bandwidth allocation is received from the orchestration agent.
Only if the new configuration is available in switches, we release the current configuration to transition to the new one accordingly so that we can hide the deletion-creation interval.
In addition, the information exchange between the transport manager and orchestration agent is support through the VR-T (virtual resource - transport) interface and the VR interface. The association of users and slices in the transport network are identified by using their source and destination IP addresses.

\subsection{Computing Manager}
The computing manager is designed to dynamically allocate computing resources, e.g., the number of CUDA \textit{threads}, in the CUDA-based GPU computing platform.  
In the CUDA programming model, an application can launch multiple \textit{kernels}, where every kernel can be concurrently executed by massive CUDA threads~\cite{nvidia2011nvidia}.
The number of threads required by a kernel is specified in its execution configuration syntax.
The execution of these kernels in the kernel space follows the order of their callings in the user space.
With the multiple-processes service (MPS), multiple applications or processes can share the GPU simultaneously.
However, the resource scheduling strategies of user applications are nontransparent and not revealed by NVIDIA.
As a result, the resource usage of user applications can not be effectively controlled.

To address this issue, we develop a kernel-split mechanism to control the GPU computing resources by managing the maximum concurrent number of threads occupied by every user application. The kernel-split mechanism splits a kernel that requests a large number of threads into multiple small and consecutive kernels with a specific number of threads.
We heavily modify the kernels of user applications to dynamic split the kernels according to the user's virtual resources at runtime. Since the execution of kernels are in-order and consecutive, the number of threads occupied by a user application always less than its virtual resources.
We develop the VR-C (virtual resource - computing) interface in the computing manager for exchanging information with the orchestration agent. The association between a mobile user and the network slice is identified by the IP address.

\subsection{System Monitor}
The system monitor is designed to collect information of network state, e.g., traffic load and slice performance, by using a dataset. The database also records the user-slice association based on the users' IMSIs and IP addresses. The system monitor uses the VR interface to communicate with radio, transport and computing manager. 

The RC (resource coordination) interface is developed to allow the central performance coordinator to communicate with orchestration agents and system monitors through the RC-L (resource coordination - learning) and RC-M (radio coordination - monitoring), respectively. The SR (slice request) interface is developed to enable the slice tenants to request and configure their network slices. For example, slice tenants can make and modify their service-level agreements (SLAs) with network operator.
The SLAs will be enforced during the resource orchestrations.

\begin{table*}[!t]
    \small
	\centering
	\begin{tabular}{|c|c|c|}
		\hline
	   \textbf{Component}     &  \textbf{Hardware}  &  \textbf{Software} \\ \hline
   	   UEs     & 4x Samsung smartphones with band selection capability &  Android 7.0\\ \hline
	   eNodeBs     & 2x Intel i5 Computer with low-latency kernel 3.19 &  OpenAirInterface (OAI)~\cite{OAI}\\ \hline
	   RF Front-End     & 2x Ettus USRP B210  &  N/A\\ \hline
	   Transport     &   6x OpenFlow 1.3 Ruckus switches &  OpenDayLight-Boron~\cite{medved2014opendaylight}\\ \hline
	   Core Network     & Intel i7 desktop computer &  openair-cn~\cite{openaircn} \\ \hline
	   Edge Servers     & 2x NVIDIA GEFORCE GTX 1080Ti  &  CUDA 9.0~\cite{nvidia2011nvidia} \\ \hline
	\end{tabular}
	\caption{Details of the Prototype}\label{tbl:hw_detail}
\vspace{-0.15in}
\end{table*}

\begin{figure}[t]
	\centering
	\includegraphics[width=3.4in]{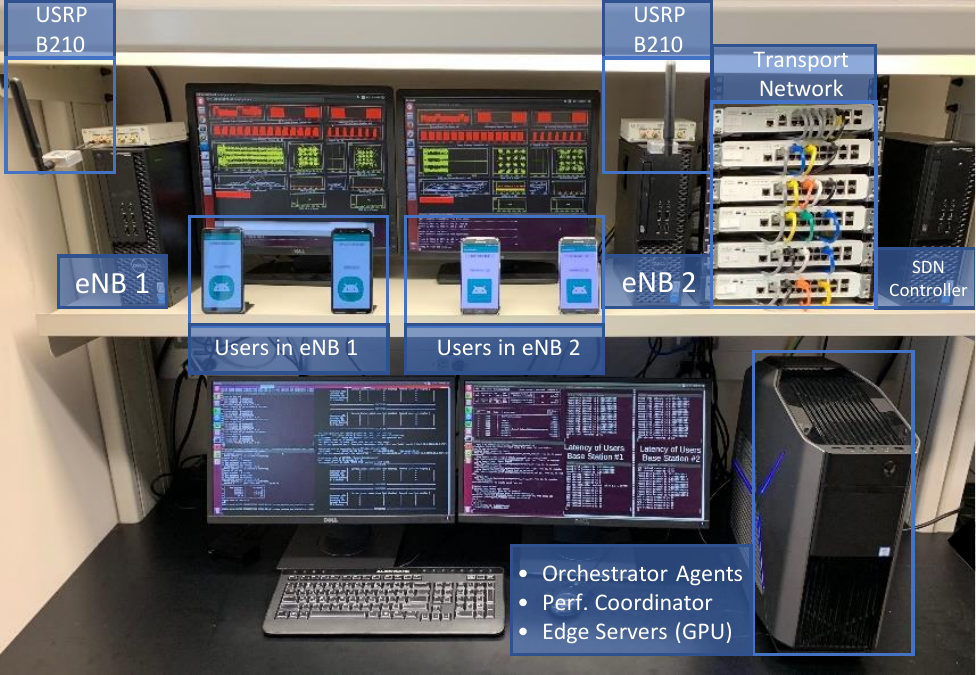}
	\caption{\small The overview of prototype.}
	\label{fig:testbed_overview}
	\vspace{-0.15in}
\end{figure}

\section{System Implementation}
\label{sec:implementation}
\subsection{Hardware Details}
We develop a prototype of the EdgeSlice system as depicted in Fig.~\ref{fig:testbed_overview}. 
It is composed of a RAN with 2 eNodeBs, a transport network with 6 OpenFlow switches, a core network, and 2 edge servers with CUDA GPUs.
The details of hardware are summarized in Table~\ref{tbl:hw_detail}.
To eliminate the co-channel interference, eNodeBs are operating at different frequency bands, i.e., LTE Band 7 and Band 38.
We configure the band selection option on smartphones so that the users in eNodeB 1 and 2 can only search for band 7 and band 38, respectively.

In the prototype, there are 2 RAs, 2 slices and 4 mobile users (1 user per slice per RA), where a RA is the set of an eNodeB, an edge server and a transport link.
The orchestration agents and performance coordinator are implemented in the core network (Alienware R7 desktop) with Python 3.5.
The optimization toolbox used in the performance coordinator is CVXPY 1.0~\cite{cvxpy}.
The radio manager is deployed in every eNodeB.
The transport manager is deployed on an individual desktop computer.
The computing manager is implemented on the edge server for every RA.
Both eNodeBs are with 5MHz (25 PRBs) wireless bandwidth.
The total bandwidth between an eNodeB and its corresponding edge server is 80Mbps.
The total amount of the computing resource for each RA is 51200 CUDA threads.

We implement orchestration agents with Tensorflow 1.10~\cite{abadi2016tensorflow}.
We use a 2-layer fully-connected neural network in both actor and critic networks.
Both layers adopt Leaky Recifier~\cite{goodfellow2016deep} activation functions with 128 neurons.
In the output layer, we use $sigmoid$~\cite{goodfellow2016deep} as the activation function.
On training orchestration agents, we conduct extensive and empirical tunings on the hyper-parameters.
We randomly generate $z_{i,j}-y_{i,j} $ between 0 and $R^{tot}_j$ to train the agents under different coordinating information.
The parameter $ \beta =20$ to have sufficient weight on enforcing the total orchestrated resources constraint~(\ref{const2}). 
The learning rates of both actor and critic networks are 0.001. The batch size is 512.
The total training step is 1E6.
The discounted factor for cumulative reward is $\gamma=0.99$.
We add the decaying Gaussian noise on actions during the training phase for balancing the exploitation and exploration.
The noise starts from $\mathcal{N}(0,1)$ and decays with factor 0.9999 per update step.

\subsection{Simulated Network Environment}
\begin{figure}[t]
	\centering
	\includegraphics[width=3.2in]{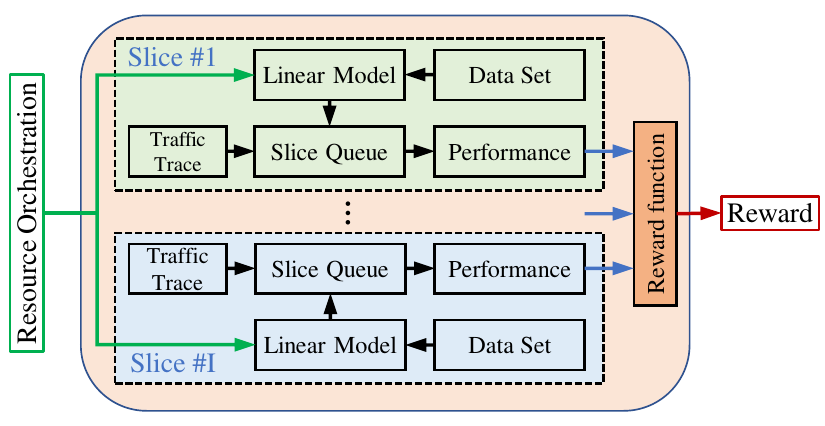}
	\caption{\small The simulated network environment.}
	\label{fig:network_env}
	\vspace{-0.15in}
\end{figure}

The orchestration agents are trained offline by using a simulated network environment as shown in Fig.~\ref{fig:network_env}.
In the environment, we implement a first-in first-out (FIFO) queue for services in individual network slices, and the performance function of each slice can be customized.
In each time interval, the traffic, i.e., service tasks, in the network slices is generated according to real network traffic traces~\cite{trace_data}. The service time of each task is determined by the end-to-end resource orchestration. 

With the simulated network environment, we generate the training dataset by traversing all possible orchestration actions using the grid search method for radio, transport and computing resources, respectively.
Due to the large number of orchestration actions, we conduct the experiments with resource granularity $10\%$ for all the resources, which means the dataset only contains discrete orchestration actions.
During the training of agents, it may produce orchestration actions that are not contained in the training dataset. To solve this problem, we build a linear regression model with \emph{scikit-learn}~\cite{scikit-learn} tool to approximate the correlations between orchestration actions and the slice performance.
Given a resource orchestration action such as $[12, 38, 22]\%$, we use adjacent orchestration actions in the dataset, e.g., $[10, 30, 20]\%$ and $[10, 40, 20]\%$, to fit the linear model.
Once the linear model is fitted, it makes the prediction for the service time under the orchestration action. The service time determines the traffic departure in service queues.
At the end of each time interval, the reward is derived based on the performances of all network slices and the design of reward function in Eq.~\ref{eq:reward_function}.  


\section{Performance Evaluation}
\label{sec:evaluation}
In this section, we evaluate the performance of EdgeSlice with both prototype experiments and network simulations.
At each time interval, the $i$th slice on the $j$th RA reports its performance to orchestration agent according to 
$\mathbf{U}_{i,j}^{(t)} = - ( l_{i,j}^{(t)} )^\alpha, \forall i \in \mathcal{I}, j \in \mathcal{J}, t \in  \mathcal{T}$,
where $\alpha = 2$ and $l_{i,j}^{(t)}$ is the queue length. Note that the performance function is defined to evaluate whether EdgeSlice can learn the optimal resource orchestration policy. In other words, neither the performance coordinator or orchestration agent know the closed-form expression of the performance function. Besides, various performance functions are evaluated in simulations.
The performance requirements of slices are defined as $\mathbf{U}_{i}^{\min} = -50, \forall i \in \mathcal{I}$ and $\rho=1.0$~\cite{hong2017linear}.

\subsection{Mobile Application}
To evaluate the system performance, we develop a mobile application which offloads computation tasks to the edge/cloud servers. Here, the computation tasks are the video analysis based on the YOLO object detection framework~\cite{redmon2016you}.
The basic procedures of these applications are:
1) a user sends a video frame with a specific resolution to server and waits to receive the processed results;
2) the server receives the frame from the user and executes the YOLO algorithm with a specific computation model to analyze the frame;
3) the server sends the analysis results back to the user.
The mobile application can use different frame resolutions, e.g., 100x100, 300x300 to 500x500, and select computation models, e.g., YOLO 320x320, YOLO 416x416 to YOLO 608x608. Here, the application with a higher frame resolution has heavier transmission traffic, and the application with a larger computation model requires a more intensive computation workload.


\subsection{Comparison Algorithms}
In the performance evaluation, we compare the EdgeSlice resource orchestration with the following algorithms:

\textbf{Traffic-Aware Resource Orchestration (TARO)}: TARO is the baseline algorithm in which all the resources are proportionally shared by slices according to the current queue length. In other words, $\mathbf{x}_{i,j}^{(t)} = R^{tot}_j \cdot l_{i,j}^{(t)} / \sum \nolimits_{i \in \mathcal{I}}{l_{i,j}^{(t)}}, \forall j \in \mathcal{J}$. This sharing scheme applies to all the RAs in the system.

\textbf{EdgeSlice-Non-Traffic (EdgeSlice-NT)}: EdgeSlice-NT is a simplified version of EdgeSlice in which the orchestration agent manages resources only based on the coordination information from the performance coordinator. Therefore, the state space of the orchestration agent of EdgeSlice-NT is $\mathbf{s}_t = [ z_{i,j}-y_{i,j}, \forall i \in \mathcal{I} ]$. In other words, EdgeSlice-NT does not use queue length of network slices as the state in the DRL model. By comparing EdgeSlice and EdgeSlice-NT, we can evaluate the impact of the state space design, i.e. whether including traffic load or not, on the performance of network slices.  





\subsection{Experimental Results}
Here, we present the experimental results and evaluate the performance of the EdgeSlice system from different angles.
In the experiment, there are 2 slices, 2 RAs and 3 types of resources.
The mobile application in the first slice uses 500x500 frame resolution and selects YOLO 320x320 as the computation model.
This application represents the type of applications that have heavy transmission traffic load and moderate computation workload.
The mobile application in the second slice uses 100x100 frame resolution and selects YOLO 608x608 as the computation model.
This application represents the type of applications that have light transmission traffic load and intensive computation workload.

In the experiments, the time interval $t$ is 1 second and the time period $\mathcal{T}$ is composed of 10 time intervals.
During the time intervals, the task arrival of network slices follow the Poisson process with average arrival rate\footnote{The slice traffic is normalized based on the hardware capability of the prototype such as bandwidth and GPU on accommodating the mobile applications.} 10.

\subsubsection{Convergence}
\begin{figure}[!t]
\centering
\includegraphics[width=3.3in]{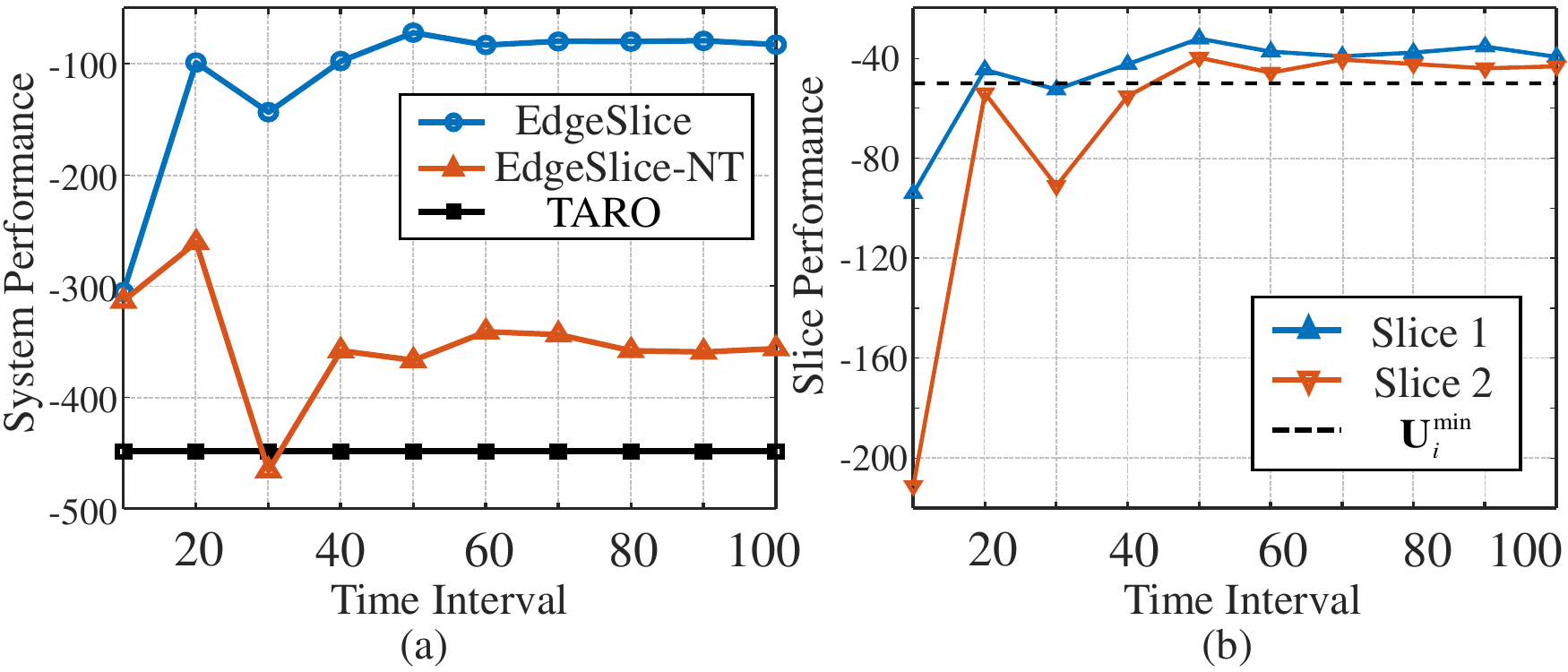}
\vspace{-0.07in}
\caption{\small The convergence of algorithms: (a) system performance vs. time interval; (b) slice performance vs. time interval. }
\label{fig:testbed_comb_convergence}
\vspace{-0.15in}
\end{figure}

In the EdgeSlice system, the performance coordinator coordinates multiple orchestration agents via the coordinating information [$z_{i,j} - y_{i,j}, \forall i \in \mathcal{I}$]. We first evaluate how fast the interaction between the coordinator and orchestration agents can converge. As depicted in Fig.~\ref{fig:testbed_comb_convergence} (a), both EdgeSlice and EdgeSlice-NT are able to converge after several time periods.
This result also reveals that orchestration agents can effectively orchestrate resources to slices under different coordinating information.
EdgeSlice obtains 3.69x and 2.74x improvement on the system performance as compared to TARO and EdgeSlice-NT, respectively.
The performance gain over TARO proves that EdgeSlice can effectively learn the optimal resource orchestration policy based on the current network state and the coordinating information. 
The performance gain over EdgeSlice-NT indicates that observing the traffic load of slices by orchestration agents can significantly improve the system performance.
In addition, as shown in Fig.~\ref{fig:testbed_comb_convergence} (b), the EdgeSlice system ensures that both network slices meet their minimum performance requirements.


\begin{figure}[!t]
\centering
\includegraphics[width=3.4in]{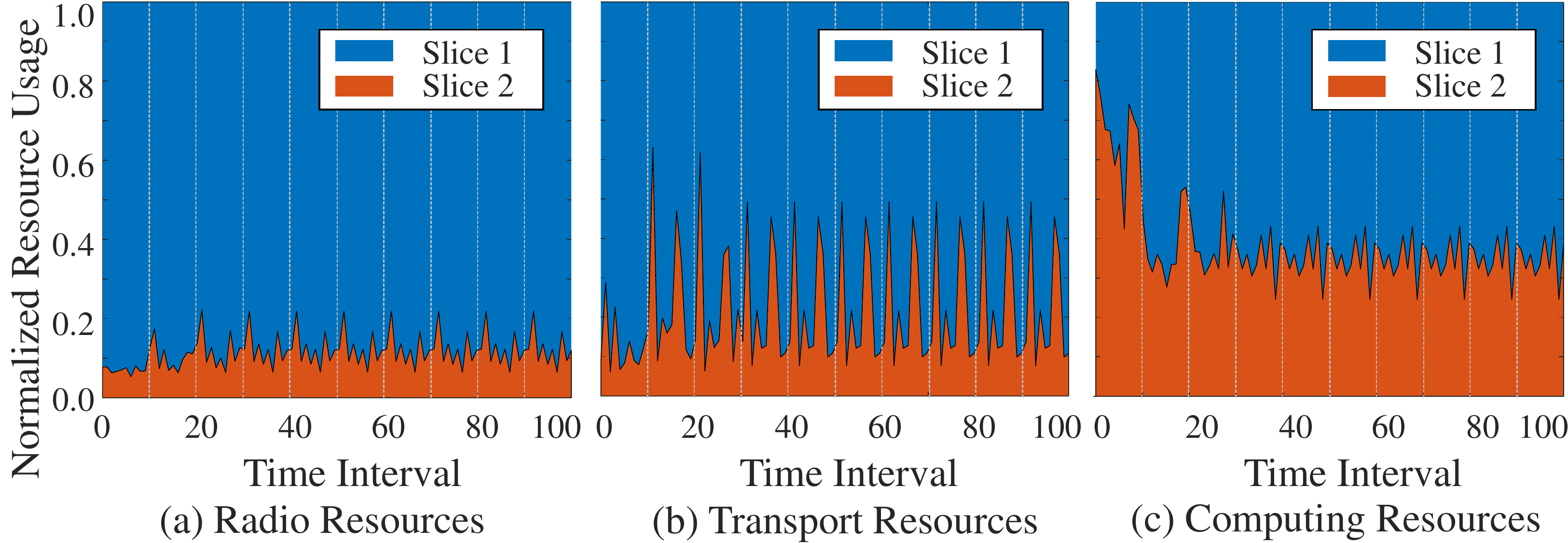}
\vspace{-0.07in}
\caption{\small The multiple resource orchestrations of EdgeSlice: (a) radio resource; (b) transport resource; (c) computing resource.}
\label{fig:testbed_allocation}
\vspace{-0.2in}
\end{figure}

\begin{figure*}[!t]
\centering
\includegraphics[width=7in]{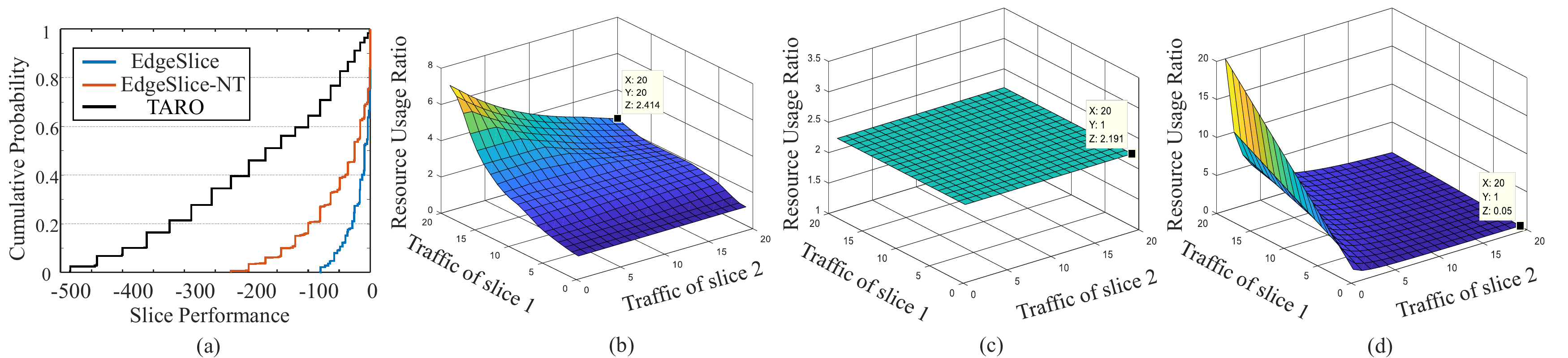}
\vspace{-0.07in}
\caption{\small The performance of orchestration agents: (a) the CDF of performance; (b) $\eta_1 / \eta_2$ vs. slice traffic under EdgeSlice; (c) $\eta_1 / \eta_2$ vs. slice traffic under EdgeSlice-NT; (d) $\eta_1 / \eta_2$ vs. slice traffic under TARO.}
\label{fig:testbed_comb_agent}
\vspace{-0.2in}
\end{figure*}

Fig.~\ref{fig:testbed_allocation} shows the normalized usage of multiple resources, i.e., radio, transport and computing resources, with the EdgeSlice system. 
In the experiments, slice 1 has a higher demand of radio and transport resources and a lower demand of computing resources than slice 2 does.
Hence, we observe that EdgeSlice allocates more radio and transport resources to slice 1 (blue area). Since slice 2 serves compute-intensive applications, it requires more computing resources. Therefore, in the beginning, slice 2 is allocated more computing resources. Later, EdgeSlice observes that the performance requirement of slice 1 cannot be met although it is allocated almost all the radio and transport resources. Thus, EdgeSlice starts to allocate more computing resources to slice 1 and then the resource orchestration converges.
Moreover, we observe the resources orchestrations becomes stable after 6 interactions, which corresponds to the observations in Fig.~\ref{fig:testbed_comb_convergence} (a).

\subsubsection{Resource Orchestration}
We evaluate the orchestration agent without any central coordination to understand its resource orchestration policy.
Fig.~\ref{fig:testbed_comb_agent} (a) depicts the cumulative distribution function (CDF) of the slice performance under randomly generated slice traffic loads.
We can see that EdgeSlice substantially outperforms both TARO and EdgeSlice-NT in terms of the slice performance.
For example, 80\% of the slice performance is larger than -30 using EdgeSlice while it is only 11\% and 55\% using TARO and EdgeSlice-NT, respectively.
The performance difference between EdgeSlice and EdgeSlice-NT is smaller than that shows in Fig.~\ref{fig:testbed_comb_convergence} (a). The reason is that the performance deficiency of the orchestration agent in EdgeSlice-NT accumulates during the iterative interactions between the agents and the coordinator. 

Fig.~\ref{fig:testbed_comb_agent} (b)-(d) show the average resource usage ratio between slice 1 and slice 2 obtained by using EdgeSlice under different traffic loads. 
The average resource usage of a slice is calculated as $\eta_i = \sum\nolimits_{k \in \mathcal{K}} x_{i,j,k}/ r_{j,k}^{tot}$.
It can be observed that EdgeSlice allocates resources to slices based on both traffic load and the application's resource needs in different domains. For example, when traffic loads of slice 1 and slice 2 are 20 and 5, respectively, the average resource usage ratio is about 5. This example shows the traffic-awareness of EdgeSlice. 
Since the orchestration agent in EdgeSlice-NT does not learn the slice traffic load in the resource orchestration, the resource usage ratio is a constant as shown in Fig.~\ref{fig:testbed_comb_agent} (c).
TARO allocates resources purely based on the slice traffic and is not aware of the actual resource needs from each domain. The resource usage ratio with TARO is shown in Fig.~\ref{fig:testbed_comb_agent} (d). The comparison between EdgeSlice and TARO shows that EdgeSlice is aware of the multi-domain resource needs of an application. 
These results validate that orchestration agents of EdgeSlice are able to autonomously orchestrate end-to-end resources under varying slice traffic.

\subsection{Simulation Results}
We set up network simulations to evaluate EdgeSlice in terms of scalability and working with different training techniques and performance functions. In the simulation, there are 5 slices, 10 RAs, and 3 types of resources.
The applications served by the network slices randomly select the frame resolutions, e.g., 100x100, 300x300, or 500x500, and computation models, e.g., 320x320, 416x416, 608x608. 
%
We use the network trace from an Italy mobile network over the Province of Trento~\cite{trace_data} to generate the traffic in network slices.
The network trace contains 154.8M entries with a minimum 10 minutes time interval collected in December 2013.
Each entry includes the counts of phone calls, SMS, Internet traffic, and the geographic square area id.
We obtain the average calling traffic in 24 hours under different geographic areas, and use them for the traffic of network slices.
In the simulation, the time interval $t$ is 1 hour and the time period $\mathcal{T}$ is composed of 24 time intervals.

\subsubsection{Scalability of EdgeSlice}
\begin{figure}[!t]
\vspace{-0.1in}
\centering
\includegraphics[width=3.3in]{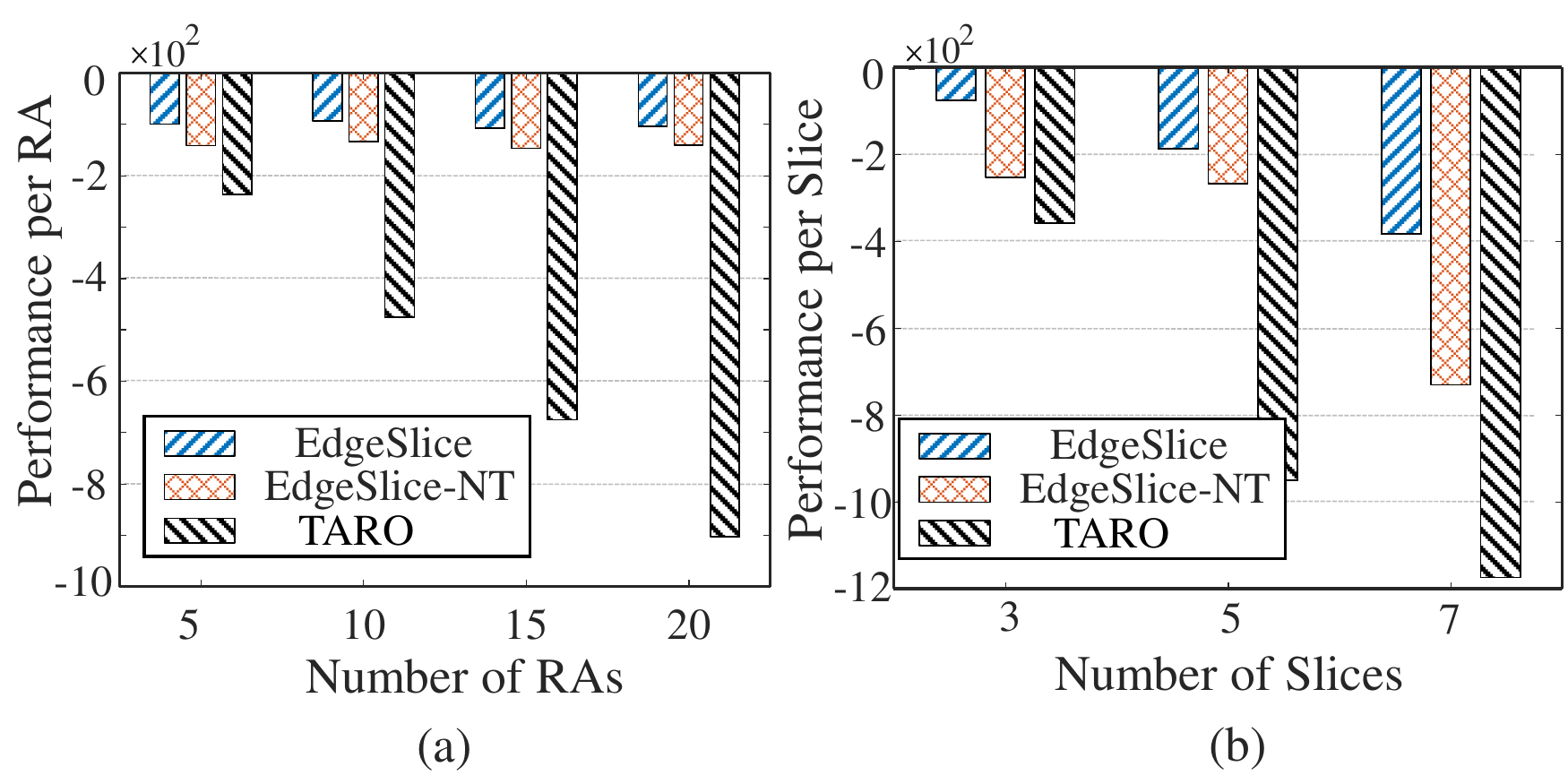}
\vspace{-0.1in}
\caption{\small The scalability of EdgeSlice: (a) performance per RA vs. the number of RAs; (b) performance per slice vs. the number of slices.}
\label{fig:simulation_scalablity}
\vspace{-0.2in}
\end{figure}

We evaluate the scalability of EdgeSlice by varying the number of slices and RAs.
As shown in Fig.~\ref{fig:simulation_scalablity} (a), both EdgeSlice and EdgeSlice-NT maintain similar performance per RA as the number of RAs increases, while the performance per RA of TARO decreases substantially.
This result indicates the EdgeSlice agents learn much superior resource orchestration policy than TARO in each RA.
Besides, EdgeSlice is capable of scaling to large network sizes without noticeably sacrificing system performance.
Fig.~\ref{fig:simulation_scalablity} (b) shows the performance per slice versus different number of network slices. As the number of slices increases, the system performance decrease because the resource demand is higher and the average allocated resources of slices are reduced. Nevertheless, EdgeSlice is still able to obtain a better performance than the others.
These results validate the scalability of the EdgeSlice system.

\subsubsection{Training Techniques of Agents}
\begin{figure}[!t]
\centering
\includegraphics[width=3.3in]{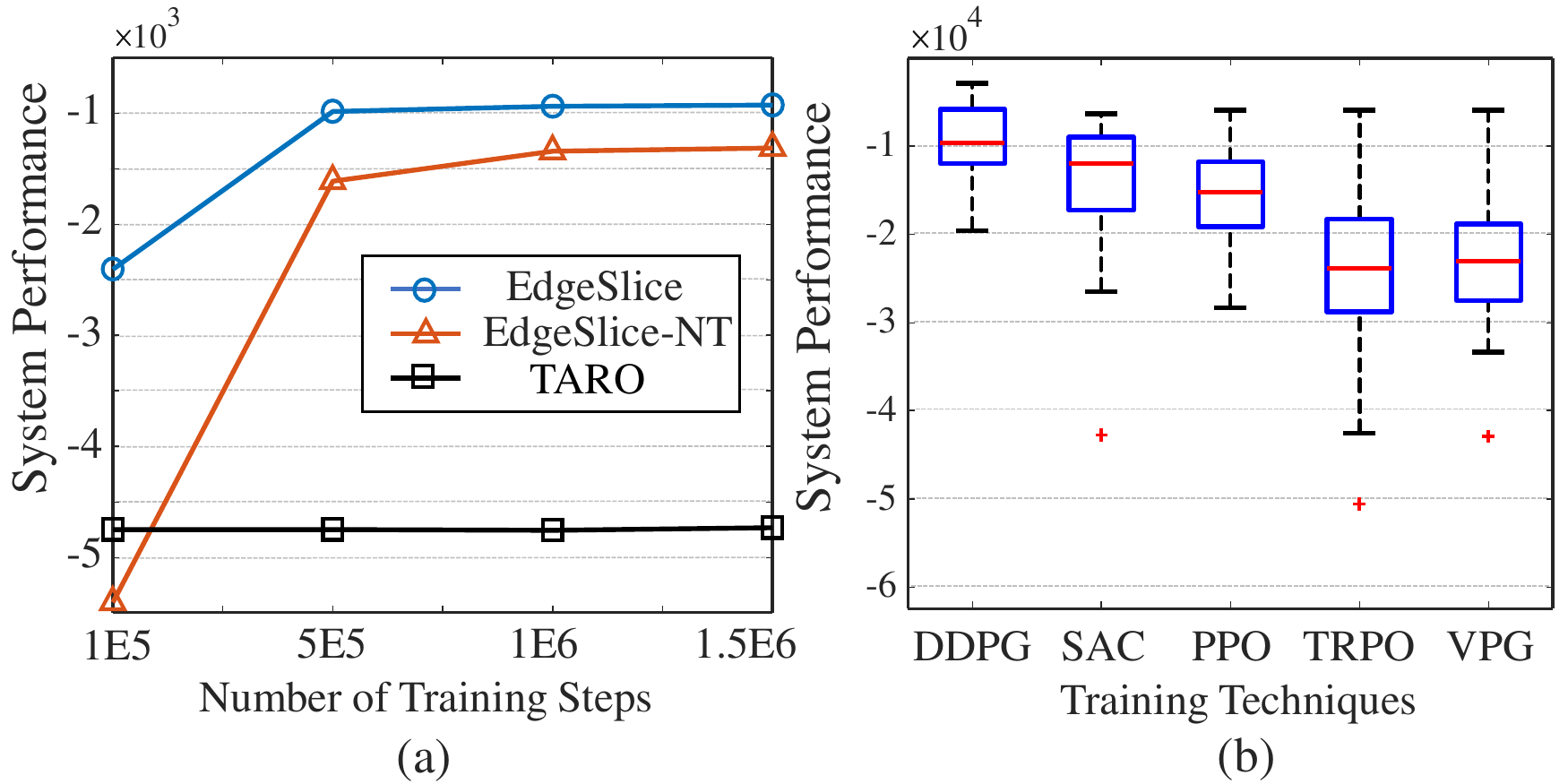}
\vspace{-0.1in}
\caption{\small The impact of training techniques: (a) system performance vs. the number of training steps of orchestration agents; (b) system performance vs. various training techniques. }
\label{fig:simulation_training_technique}
\vspace{-0.1in}
\end{figure}

We study the impact of various techniques on training the orchestration agents in the EdgeSlice system.
As depicted in Fig.~\ref{fig:simulation_training_technique} (a), the system performance drops remarkably when the training steps of agent is insufficient such as 1E5.
In general, a learning-based agent with a large number of training steps has better performance than that with a small number of training steps.
We can see that the performance of EdgeSlice and EdgeSlice-NT can be worse than that of TARO if the number of training steps is 1E5 or less. This means that if the agent is not well trained, it could lead to very poor performance. 
Moreover, various techniques, e.g., SAC~\cite{haarnoja2018soft}, PPO~\cite{schulman2017proximal}, TRPO~\cite{schulman2015trust}, and VPG~\cite{sutton2000policy}, have been proposed to improve the performance of agents.
We evaluate the system performance of EdgeSlice under different training techniques as shown in Fig.~\ref{fig:simulation_training_technique} (b).
The training setting and hyper-parameters are the same as mentioned in Sec.~\ref{sec:implementation}.
The orchestration agent trained using DDPG exhibits better performance than that trained by the other techniques. These results show the importance of the training techniques in developing the EdgeSlice system. 

\begin{figure}[!t]
\vspace{-0.1in}
\centering
\includegraphics[width=3.3in]{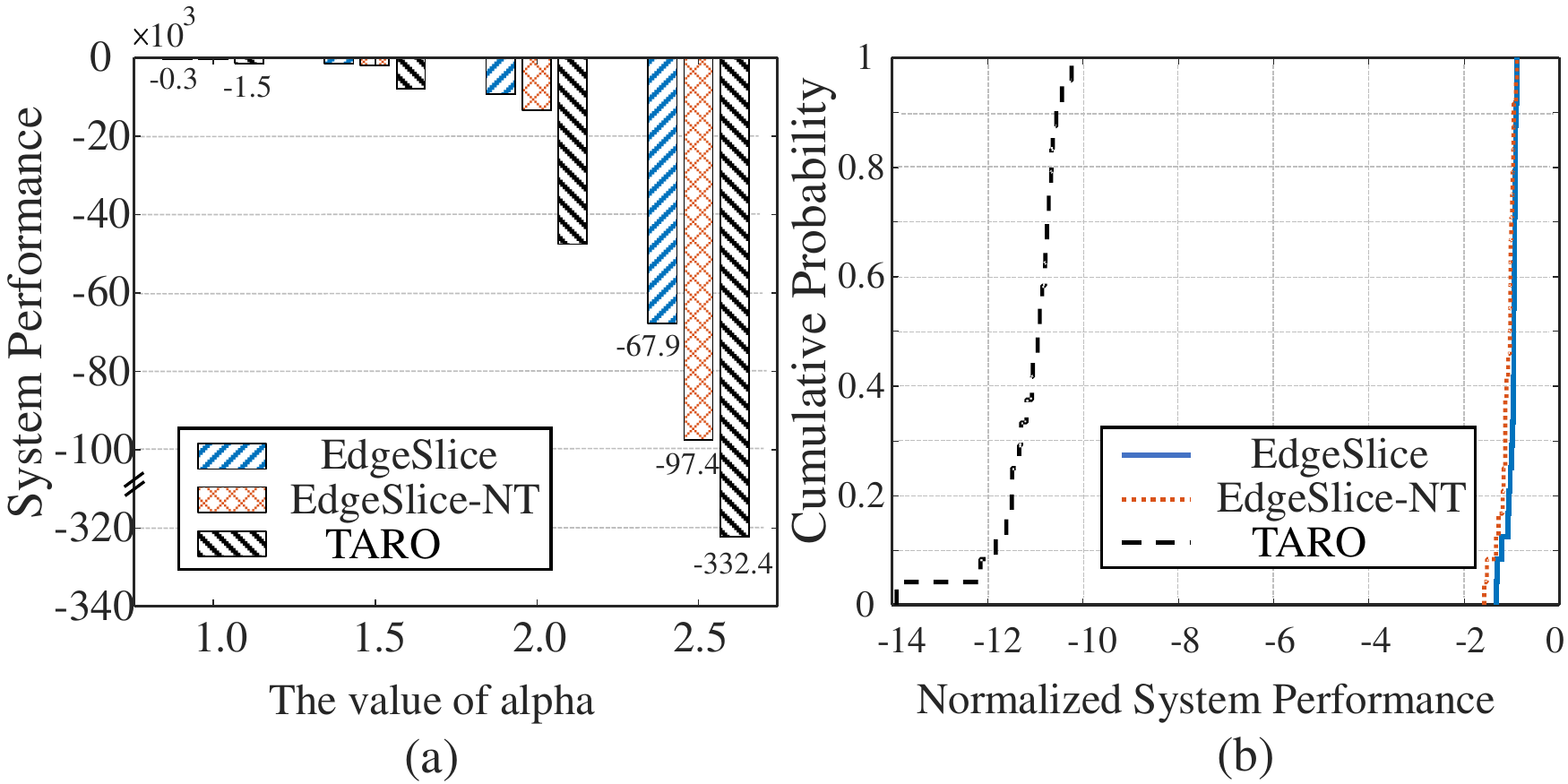}
\vspace{-0.1in}
\caption{\small The compatibility of EdgeSlice: (a) system performance vs. the value of $\alpha$, (b) CDF of normalized system performance.}
\label{fig:simulation_performance_function}
\vspace{-0.2in}
\end{figure}

\subsubsection{Handling different performance functions}
We evaluate the performance of EdgeSlice under different performance functions of network slices.
As shown in Fig.~\ref{fig:simulation_performance_function} (a), we vary the value of $\alpha$ in the performance function. The large $\alpha$ indicates slice reports worse performance under the same queue length. The EdgeSlice outperforms the others for all conditions, which implies EdgeSlice can automatically learn superior resource orchestration policy under varying performance functions.
Furthermore, we define another performance function as the negative service time of slice users without considering traffic in slice queue. As shown in Fig.~\ref{fig:simulation_performance_function} (b), EdgeSlice and EdgeSlice-NT achieve almost the same system performance.
Because we intentionally eliminate the impact of slice traffic on the slice performance function. As a result, the network state, i.e., queue length, observed by EdgeSlice is not helpful on learning the correlations.
In contrast, the performance of TARO is much worse. These results indicate that when the performance function is less dependent on the network state, learning-based EdgeSlice and EdgeSlice-NT still have much performance gain over TARO. 
These results verify the capability of EdgeSlice on handling various performance functions of slices.

\section{Related Work}
This work relates to resource management in network slicing and deep reinforcement learning for networking problems. 

\textbf{Resource Management in Network Slicing:}
The resource management problem in network slicing has been extensively studied with the goal to maximize the system performance.
Caballero \emph{et al.}~\cite{caballero2019network} constructed a network slicing game in which tenants are selfish to maximize its own performance. The authors proved that this game with such strategic behavior converges to a Nash equilibrium for elastic traffic.
Halabian \emph{et al.}~\cite{halabian2019distributed} showed that non-collaborative slices in the system compromise the fairness performance when maximizing the overall system performance and proposed a distributed solution by extending the Dominant Resource Fairness (DRF) framework. 
To exploit the statistical multiplexing gain of slices, Sciancalepore \emph{et al.}~\cite{sciancalepore2019storns} designed STORNS that optimizes the admission control of slices with considering per slice SLA requirement by leveraging stochastic geometry theory.
Salvat \emph{et al.}~\cite{salvat2018overbooking} developed an end-to-end resource orchestration system, formulated an orchestration problem to maximize the revenue in network slicing, and proposed an optimal Benders decomposition method and a heuristic method.
Foukas \emph{et al.}~\cite{foukas2017orion} developed an efficient RAN slicing system that enables the dynamic and real-time virtualization of base stations and slices customization to meet slices' service demands.
However, the fundamental assumption of these works is that the resource demands of slices and their performance functions are known as closed-form mathematical expressions to network operator. In contrast, the EdgeSlice system proposed in the paper enables a model-free resource orchestration solution.

\textbf{Deep Reinforcement Learning (DRL) in Networking:}
Machine learning techniques such as deep learning and reinforcement learning gain significant popularity in solving resource management problems in mobile network for coping the complicated network dynamics.
Mao \emph{et al.}~\cite{mao2016resource} designed DeepRM with the DQN technique to optimize the admission control and resource orchestration of users. They obtained a considerable reduction on the average slowdown of user tasks as compared to heuristic solutions.
Xu \emph{et al.}~\cite{xu2018experience} utilized the state-of-the-art DDPG technique to solve the traffic engineering (TE) networking problem, i.e., allocating the bandwidth of network links, and obtained significant end-to-end latency reduction and performance improvement under the unknown performance function.
Bega \emph{et al.}~\cite{bega2019deepcog} proposed DeepCog with deep learning techniques to forecast the network capacity within an individual slice and achieve the balance between resource over-provisioning and service request violations. 
%
Yang \emph{et al.}~\cite{yang2019miras} proposed an adaptive reinforcement learning based approach for microservice workflow system that enables model-free resource allocation and improves the response time of microservices.
However, these existing works advocate the centralization of resource management by using a central agent. Although their solutions may be applied to perform the resource orchestration, the centralized resource orchestration is highly complex for wireless edge computing networks. Different from these methods, the EdgeSlice system enables a decentralized resource orchestration. 

\vspace{-0.07in}
\section{Conclusion}
In this paper, we have designed EdgeSlice, a new decentralized resource orchestration system, to automate dynamic network slicing in wireless edge computing networks. To realize EdgeSlice, we have developed a novel decentralized deep reinforcement learning method which consists of a central performance coordinator and multiple orchestration agents. The orchestration agent learns the optimal resource orchestration policy for network slicing under the coordination of the central performance coordinator. We have also designed new radio, transport and computing resource manager that enable dynamic configuration of end-to-end resources at runtime.
We have developed a prototype of EdgeSlice with OpenAirInterface (OAI) in radio access network, OpenDayLight (ODL) in transport network, and CUDA GPU computing in edge/cloud servers.
The performance of EdgeSlice has been validated through both prototype implementation and network simulations. 

\vspace{-0.1in}
\section*{Acknowledgement}
This work is partially supported by the US National Science Foundation under Grant No. 1731675, No. 1810174, and No. 1910844.
\vspace{-0.1in}
\bibliographystyle{IEEEtran}
\vspace{-0.07in}
\bibliography{ref/reference}

\begin{thebibliography}{10}
\providecommand{\url}[1]{#1}
\csname url@samestyle\endcsname
\providecommand{\newblock}{\relax}
\providecommand{\bibinfo}[2]{#2}
\providecommand{\BIBentrySTDinterwordspacing}{\spaceskip=0pt\relax}
\providecommand{\BIBentryALTinterwordstretchfactor}{4}
\providecommand{\BIBentryALTinterwordspacing}{\spaceskip=\fontdimen2\font plus
\BIBentryALTinterwordstretchfactor\fontdimen3\font minus
  \fontdimen4\font\relax}
\providecommand{\BIBforeignlanguage}[2]{{%
\expandafter\ifx\csname l@#1\endcsname\relax
\typeout{** WARNING: IEEEtran.bst: No hyphenation pattern has been}%
\typeout{** loaded for the language `#1'. Using the pattern for}%
\typeout{** the default language instead.}%
\else
\language=\csname l@#1\endcsname
\fi
#2}}
\providecommand{\BIBdecl}{\relax}
\BIBdecl

\bibitem{agiwal2016next}
M.~Agiwal, A.~Roy, and N.~Saxena, ``Next generation {5G} wireless networks: A
  comprehensive survey,'' \emph{IEEE Communications Surveys \& Tutorials},
  vol.~18, no.~3, pp. 1617--1655, 2016.

\bibitem{foukas2017network}
X.~Foukas, G.~Patounas, A.~Elmokashfi, and M.~K. Marina, ``Network slicing in
  {5G}: Survey and challenges,'' \emph{IEEE Communications Magazine}, vol.~55,
  no.~5, pp. 94--100, 2017.

\bibitem{ordonez2017network}
J.~Ordonez-Lucena, P.~Ameigeiras \emph{et~al.}, ``{Network slicing for 5G with
  SDN/NFV: Concepts, architectures, and challenges},'' \emph{IEEE
  Communications Magazine}, vol.~55, no.~5, pp. 80--87, 2017.

\bibitem{afolabi2018network}
I.~Afolabi, T.~Taleb \emph{et~al.}, ``Network slicing and softwarization: A
  survey on principles, enabling technologies, and solutions,'' \emph{IEEE
  Communications Surveys \& Tutorials}, vol.~20, no.~3, pp. 2429--2453, 2018.

\bibitem{foukas2017orion}
X.~Foukas, M.~K. Marina, and K.~Kontovasilis, ``Orion: {RAN} slicing for a
  flexible and cost-effective multi-service mobile network architecture,'' in
  \emph{ACM MobiCom}, 2017, pp. 127--140.

\bibitem{rost2017network}
P.~Rost, C.~Mannweiler \emph{et~al.}, ``Network slicing to enable scalability
  and flexibility in {5G} mobile networks,'' \emph{IEEE Communications
  magazine}, vol.~55, no.~5, pp. 72--79, 2017.

\bibitem{marquez2018should}
C.~Marquez, M.~Gramaglia, M.~Fiore, A.~Banchs, and X.~Costa-Perez, ``How should
  {I} slice my network?: A multi-service empirical evaluation of resource
  sharing efficiency,'' in \emph{MobiCom}.\hskip 1em plus 0.5em minus
  0.4em\relax ACM, 2018, pp. 191--206.

\bibitem{caballero2019network}
P.~Caballero, A.~Banchs, G.~De~Veciana, and X.~Costa-P{\'e}rez, ``Network
  slicing games: Enabling customization in multi-tenant mobile networks,''
  \emph{IEEE/ACM Transactions on Networking}, 2019.

\bibitem{halabian2019distributed}
H.~Halabian, ``Distributed resource allocation optimization in {5G} virtualized
  networks,'' \emph{IEEE Journal on Selected Areas in Communications}, vol.~37,
  no.~3, pp. 627--642, 2019.

\bibitem{kall1994stochastic}
P.~Kall \emph{et~al.}, \emph{Stochastic programming}.\hskip 1em plus 0.5em
  minus 0.4em\relax Springer, 1994.

\bibitem{salvat2018overbooking}
J.~X. Salvat \emph{et~al.}, ``Overbooking network slices through yield-driven
  end-to-end orchestration,'' in \emph{ACM CoNEXT}.\hskip 1em plus 0.5em minus
  0.4em\relax ACM, 2018, pp. 353--365.

\bibitem{mao2019learning}
H.~Mao, M.~Schwarzkopf \emph{et~al.}, ``Learning scheduling algorithms for data
  processing clusters,'' in \emph{Proceedings of the ACM Special Interest Group
  on Data Communication}.\hskip 1em plus 0.5em minus 0.4em\relax ACM, 2019, pp.
  270--288.

\bibitem{boyd2011distributed}
S.~Boyd, N.~Parikh \emph{et~al.}, ``Distributed optimization and statistical
  learning via the alternating direction method of multipliers,''
  \emph{Foundations and Trends in Machine learning}, vol.~3, no.~1, pp. 1--122,
  2011.

\bibitem{Convex2004Boyd}
S.~Boyd and L.~Vandenberghe, \emph{{Convex Optimization}}.\hskip 1em plus 0.5em
  minus 0.4em\relax Cambridge university press, 2004.

\bibitem{lillicrap2015continuous}
T.~P. Lillicrap, Hunt \emph{et~al.}, ``Continuous control with deep
  reinforcement learning,'' \emph{arXiv preprint arXiv:1509.02971}, 2015.

\bibitem{mnih2015human}
V.~Mnih, K.~Kavukcuoglu, Silver \emph{et~al.}, ``Human-level control through
  deep reinforcement learning,'' \emph{Nature}, vol. 518, no. 7540, p. 529,
  2015.

\bibitem{xu2018experience}
Z.~Xu, J.~Tang \emph{et~al.}, ``Experience-driven networking: A deep
  reinforcement learning based approach,'' in \emph{IEEE INFOCOM}.\hskip 1em
  plus 0.5em minus 0.4em\relax IEEE, 2018, pp. 1871--1879.

\bibitem{silver2018general}
D.~Silver, T.~Hubert \emph{et~al.}, ``A general reinforcement learning
  algorithm that masters chess, shogi, and go through self-play,''
  \emph{Science}, vol. 362, no. 6419, pp. 1140--1144, 2018.

\bibitem{chen2018auto}
L.~Chen, J.~Lingys \emph{et~al.}, ``{AuTO}: scaling deep reinforcement learning
  for datacenter-scale automatic traffic optimization,'' in \emph{SIGCOMM
  2018}.\hskip 1em plus 0.5em minus 0.4em\relax ACM, 2018, pp. 191--205.

\bibitem{mao2016resource}
H.~Mao, M.~Alizadeh \emph{et~al.}, ``Resource management with deep
  reinforcement learning,'' in \emph{ACM HotNets}.\hskip 1em plus 0.5em minus
  0.4em\relax ACM, 2016, pp. 50--56.

\bibitem{griffith2013policy}
S.~Griffith \emph{et~al.}, ``Policy shaping: Integrating human feedback with
  reinforcement learning,'' in \emph{Advances in neural information processing
  systems}, 2013, pp. 2625--2633.

\bibitem{konda2000actor}
V.~R. Konda and J.~N. Tsitsiklis, ``Actor-critic algorithms,'' in
  \emph{Advances in neural information processing systems}, 2000, pp.
  1008--1014.

\bibitem{bellman1966dynamic}
R.~Bellman, ``Dynamic programming,'' \emph{Science}, vol. 153, no. 3731, pp.
  34--37, 1966.

\bibitem{medved2014opendaylight}
J.~Medved, R.~Varga, A.~Tkacik, and K.~Gray, ``Opendaylight: Towards a
  model-driven {SDN} controller architecture,'' in \emph{IEEE WoWMoM
  2014}.\hskip 1em plus 0.5em minus 0.4em\relax IEEE, 2014, pp. 1--6.

\bibitem{mckeown2008openflow}
N.~McKeown, T.~Anderson \emph{et~al.}, ``Openflow: enabling innovation in
  campus networks,'' \emph{ACM SIGCOMM Computer Communication Review}, vol.~38,
  no.~2, pp. 69--74, 2008.

\bibitem{openflow_ref}
O.~S. Specification, ``Openflow switch specification version 1.5.1,
  https://www.opennetworking.org/wp-content/uploads/2014/10/openflow-switch-v1.5.1.pdf,''
  2013.

\bibitem{nvidia2011nvidia}
C.~Nvidia, ``{Nvidia CUDA C programming guide},'' \emph{Nvidia Corporation},
  vol. 120, no.~18, p.~8, 2011.

\bibitem{OAI}
{OpenAirInterface Software Alliance. OpenAirInterface repository.
  https:gitlab.eurecom.fr/oai/openairinterface5g}, 2017.

\bibitem{openaircn}
{OpenAirInterface Software Alliance. Openair-cn repository.
  https:gitlab.eurecom.fr/oai/openair-cn}, 2017.

\bibitem{cvxpy}
S.~Diamond and S.~Boyd, ``{CVXPY}: A {P}ython-embedded modeling language for
  convex optimization,'' \emph{Journal of Machine Learning Research}, vol.~17,
  no.~83, pp. 1--5, 2016.

\bibitem{abadi2016tensorflow}
M.~Abadi, P.~Barham \emph{et~al.}, ``Tensorflow: A system for large-scale
  machine learning,'' in \emph{12th $USENIX$ OSDI}, 2016, pp. 265--283.

\bibitem{goodfellow2016deep}
I.~Goodfellow \emph{et~al.}, \emph{Deep learning}.\hskip 1em plus 0.5em minus
  0.4em\relax MIT press, 2016.

\bibitem{trace_data}
T.~Italia, ``Telecommunication activity dataset,''
  \url{https://dandelion.eu/datagems/SpazioDati/telecom-sms-call-internet-tn/description/},
  2013.

\bibitem{scikit-learn}
F.~Pedregosa \emph{et~al.}, ``Scikit-learn: Machine learning in {P}ython,''
  \emph{Journal of Machine Learning Research}, vol.~12, pp. 2825--2830, 2011.

\bibitem{hong2017linear}
M.~Hong and Z.-Q. Luo, ``On the linear convergence of the alternating direction
  method of multipliers,'' \emph{Mathematical Programming}, vol. 162, no. 1-2,
  pp. 165--199, 2017.

\bibitem{redmon2016you}
J.~Redmon \emph{et~al.}, ``You only look once: Unified, real-time object
  detection,'' in \emph{IEEE CVPR}, 2016, pp. 779--788.

\bibitem{haarnoja2018soft}
T.~Haarnoja, A.~Zhou, P.~Abbeel, and S.~Levine, ``Soft actor-critic: Off-policy
  maximum entropy deep reinforcement learning with a stochastic actor,''
  \emph{arXiv preprint arXiv:1801.01290}, 2018.

\bibitem{schulman2017proximal}
J.~Schulman, F.~Wolski \emph{et~al.}, ``Proximal policy optimization
  algorithms,'' \emph{arXiv preprint arXiv:1707.06347}, 2017.

\bibitem{schulman2015trust}
J.~Schulman, S.~Levine, P.~Abbeel, M.~Jordan, and P.~Moritz, ``Trust region
  policy optimization,'' in \emph{International Conference on Machine
  Learning}, 2015, pp. 1889--1897.

\bibitem{sutton2000policy}
R.~S. Sutton, D.~A. McAllester \emph{et~al.}, ``Policy gradient methods for
  reinforcement learning with function approximation,'' in \emph{Advances in
  neural information processing systems}, 2000, pp. 1057--1063.

\bibitem{sciancalepore2019storns}
V.~Sciancalepore, M.~Di~Renzo, and X.~Costa-Perez, ``{STORNS}: Stochastic radio
  access network slicing,'' \emph{arXiv:1901.05336}, 2019.

\bibitem{bega2019deepcog}
D.~Bega, M.~Gramaglia \emph{et~al.}, ``Deepcog: Cognitive network management in
  sliced {5G} networks with deep learning,'' 2019.

\bibitem{yang2019miras}
Z.~Yang \emph{et~al.}, ``{MIRAS}: Model-based reinforcement learning for
  microservice resource allocation over scientific workflows,'' in \emph{IEEE
  ICDCS}.\hskip 1em plus 0.5em minus 0.4em\relax IEEE, 2019, pp. 122--132.

\end{thebibliography}
\vspace{-0.1in}

\end{document}